# Mathematical Modeling of Common-Pool Resources: A Comprehensive Review of Bioeconomics, Strategic Interaction, and Complex Adaptive Systems

Zebiao Li [a], Rui Liu [b], Chengyi Tu [b, *]

[a] Keyi College of Zhejiang Sci-Tech University; Shaoxing, 312369, China

[b] School of Economics and Management, Zhejiang Sci-Tech University; Hangzhou, 310018, China

[*] Corresponding author. Email: chengyitu1986@gmail.com

## Abstract

The governance of common-pool resources—resource systems characterized by high subtractability of yield and difficulty of exclusion—constitutes one of the most persistent and intricate challenges in the fields of economics, ecology, and applied mathematics. This comprehensive review delineates the historical and theoretical evolution of the mathematical frameworks developed to analyze, predict, and manage these systems. We trace the intellectual trajectory from the early, deterministic bioeconomic models of the mid-20th century, which established the fundamental tension between individual profit maximization and collective efficiency, to the contemporary era of complex coupled human-environment system models. Our analysis systematically dissects the formalization of the "Tragedy of the Commons" through the lens of classical cooperative and non-cooperative game theory, examining how the N-person Prisoner's Dilemma and Nash Equilibrium concepts provided the initial, albeit pessimistic, predictive baseline. We subsequently explore the "Ostrom Turn," which necessitated the integration of institutional realism—specifically monitoring, graduated sanctions, and communication—into formal game-theoretic structures. The review further investigates the relaxation of rationality assumptions via evolutionary game theory and behavioral economics, highlighting the destabilizing roles of prospect theory and hyperbolic discounting. Finally, we synthesize recent advances in stochastic differential equations and agent-based computational economics, which capture the critical roles of spatial heterogeneity, noise-induced regime shifts, and early warning signals of collapse. By unifying these diverse mathematical threads, this review elucidates the shifting paradigm from static optimization to dynamic resilience in the management of the commons.

# 1. Introduction

The governance of common-pool resources (CPRs) epitomizes the fundamental tension between individual incentives and collective welfare—a dilemma that has occupied scholars from Aristotle to the present. While the challenge is ancient, the rigorous mathematical formalization of CPR dynamics is a modern endeavor, catalyzed largely by industrial-era pressures on fisheries, forests, and grazing lands[1-3]. Analytically, a CPR is distinguished by two defining features: subtractability (or rivalrousness), wherein one agent's extraction diminishes the resource available to others; and non-excludability, which implies prohibitively high costs for preventing unauthorized access[4,5].

Historically, the conceptual landscape of CPRs was defined by the "Tragedy of the Commons." Popularized by Garrett Hardin in 1968, this paradigm rests on economic intuitions tracing back to William Forster Lloyd's 1833 lectures[6]. Lloyd's canonical example of the herdsman adding cattle to a shared pasture—privatizing the benefits while socializing the costs of overgrazing—established the assumption that unmanaged "open access" regimes inevitably lead to resource collapse[7]. However, Hardin's formulation was mathematically implicit, relying on narrative logic rather than formal proof[8,9].

The transition from metaphor to scientific inquiry required models capable of quantifying the interplay between resource regeneration, extraction technologies, and strategic behavior. Mathematical modeling in this domain serves three critical functions: prediction, enabling the forecasting of stock dynamics under varying pressures; efficiency assessment, specifically quantifying rent dissipation; and institutional design, providing a theoretical testbed for regulatory interventions such as quotas or sanctions[10,11].

This review traces the distinct evolution of this modeling philosophy. We begin with foundational bioeconomic models, which utilized optimal control theory to treat agents as rational, atomistic units within a deterministic environment[12]. We then examine the game-theoretic turn, which reframed the tragedy as a Nash Equilibrium and explored pathways to cooperation via the Folk Theorem. Crucially, we discuss the paradigm shift necessitated by Elinor Ostrom's empirical findings, which demonstrated that human communities frequently self-organize to avert tragedy[13]. This realization has driven contemporary research toward Complex Adaptive Systems, utilizing Coupled Human and Natural Systems frameworks, stochastic analysis, and Agent-Based Modeling to capture the co-evolution of social norms and ecological variables[14,15]. By surveying these methodologies, this review offers a comprehensive synthesis of the equations and assumptions that underpin our understanding of sustainable governance.

# 2. Deterministic Bioeconomics and the Dissipation of Rent

The foundation of CPR modeling lies in bioeconomics, a field that merges biological growth functions with economic cost-benefit analysis. Developing in the mid-20th century, largely in response to the post-war expansion of industrial fishing, bioeconomics sought to determine the "optimal" level of harvest that balances biological regeneration with economic gain[16-18]. These models primarily utilize ordinary differential equations (ODEs) to describe the trajectory of a resource stock over time[19].

## 2.1 Gordon-Schaefer Model

The theoretical foundation of modern bioeconomics is the Gordon-Schaefer model, a seminal framework established independently by Gordon and Schaefer[20-22]. By synthesizing the biological dynamics of logistic growth with an economic production function, this model provided the first unified theory of fishery exploitation and remains the definitive pedagogical baseline for analyzing the mechanics of open-access rent dissipation[23-25].

The biological dynamics of the resource are governed by the logistic growth function[26,27], originally proposed by Verhulst. Letting $B(t)$ denote the biomass at time $t$, the intrinsic rate of change in the absence of harvest is described by:

$$\frac{dB}{dt} = F(B) = rB\left(1 - \frac{B}{K}\right)$$

where $r$ represents the intrinsic growth rate and $K$ denotes the environmental carrying capacity. This formulation yields a sigmoid growth trajectory: the population exhibits exponential expansion at low densities (driven by the $rB$ term) before saturating as it approaches $K$ due to density-dependent constraints, such as resource limitation. Consequently, the maximum biological productivity — Maximum Sustainable Yield (MSY) — is achieved at $B_{MSY} = K/2$, where the instantaneous growth rate $F(B)$ is maximized.

Harvesting is incorporated as an exogenous mortality source driven by economic activity. The harvest rate, $Y(t)$, is typically modeled using the Schaefer production function, which assumes a bilinear relationship between fishing effort $E(t)$ and standing biomass $B(t)$:

$$Y(t) = qE(t)B(t)$$

where $q$ is the catchability coefficient, a technological parameter quantifying the efficiency of effort. By combining these components, the dynamic evolution of the exploited stock is $dB/dt = 0$ given by:

$$\frac{dB}{dt} = rB\left(1 - \frac{B}{K}\right) - qEB$$

This differential equation formalizes the critical feedback mechanism of the system: harvesting effort depletes biomass, which subsequently diminishes the catch per unit effort (CPUE) available for future extraction, thereby coupling economic behavior to biological constraints.

## 2.2 Equilibrium Analysis: MSY, MEY, and Open Access

The analytical utility of the Gordon-Schaefer framework[28] rests on its steady-state analysis, wherein the resource's biological growth is balanced by the harvest rate ($dB/dt = 0$). Consequently, the equilibrium biomass ($B_{eq}$) can be derived as a linear function of fishing effort ($E$): $B_{eq} = K\left(1 - \frac{qE}{r}\right)$. Substituting this relationship into the harvest function yields the Sustainable Yield curve, which exhibits a parabolic dependence on effort:

$$Y_{eq}(E) = qE \cdot K\left(1 - \frac{qE}{r}\right) = qKE - \frac{q^2 K}{r}E^2$$

From a bioeconomic perspective (see Fig. 1), the model delineates three critical reference points by converting yield into Total Revenue ($TR = pY$, where $p$ denotes a constant price) and defining a linear Total Cost function ($TC = cE$, representing the unit cost of effort). First, the Maximum Sustainable Yield (MSY) corresponds to the vertex of the yield parabola ($B = K/2$), representing the maximization of physical harvest volume. Second, the Maximum Economic Yield (MEY)[29] occurs where the differential between the TR curve and the TC line is greatest (i.e., Marginal Revenue equals Marginal Cost). This point maximizes the economic rent generated by the resource[30,31]. Notably, $E_{MEY} < E_{MSY}$, implying that economic optimality requires a more conservative effort level than biological maximization. Third, the Open Access Equilibrium (OAE) is established where Total Revenue equals Total Cost ($TR = TC$), resulting in zero economic profit. In unregulated common-pool regimes, the existence of positive rents ($TR > TC$) incentivizes entry until all economic surplus is dissipated. Formally, the open-access condition is expressed as:

$$pqEK\left(1 - \frac{qE}{r}\right) = cE$$

This equilibrium typically results in biomass levels significantly below both $B_{MSY}$ and $B_{MEY}$. Furthermore, under conditions of negligible cost ($c \to 0$) or high market valuation ($p \to \infty$), fishing effort may exceed critical thresholds, precipitating stock collapse. This dynamic provides a rigorous formalization of the "Tragedy of the Commons," demonstrating how individual rational agents can drive the complete dissipation of a resource's intrinsic value.

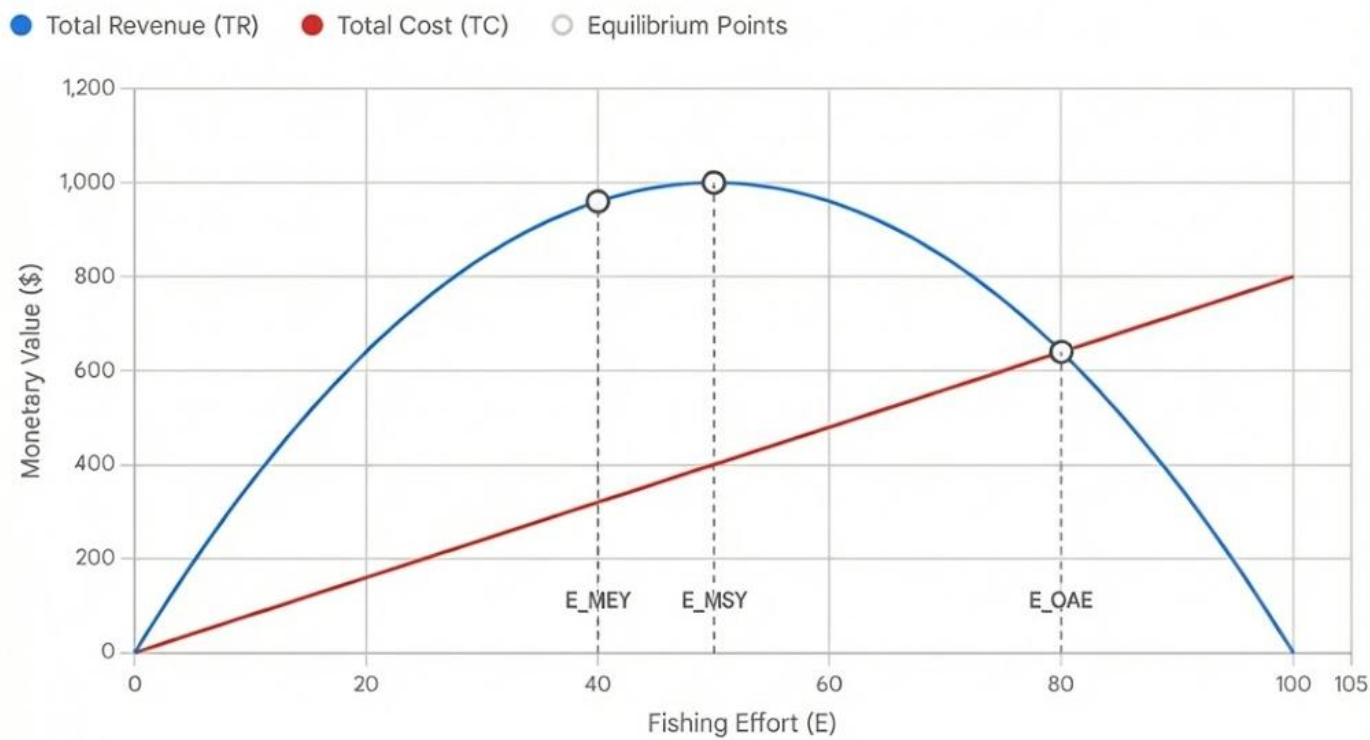

**Figure 1. Static bioeconomic equilibria and rent dissipation in the Gordon-Schaefer framework.** The parabolic arc depicts the sustainable Total Revenue (TR) curve derived from the underlying logistic growth function. The linear slope represents the Total Cost (TC) of harvesting effort. The system converges to the Open Access Equilibrium (OAE) at the intersection where $TR = TC$, signifying the complete dissipation of economic rent (zero-profit condition). The Maximum Economic Yield (MEY) is defined by the effort level that maximizes the vertical differential between revenue and cost (net economic rent), occurring at a lower intensity of exploitation than the biological Maximum Sustainable Yield (MSY).

## 2.3 Dynamic Optimization and Capital Theory

While the static Gordon-Schaefer framework provides essential equilibrium benchmarks, comprehensive resource governance requires an intertemporal perspective. A social planner must navigate the trade-off between immediate harvest and the conservation of standing stock to facilitate future growth. This challenge is formally cast as a dynamic optimization problem, treating biomass $B(t)$ as natural capital. The objective is to maximize the Net Present Value (NPV) of the resource rent over an infinite time horizon, typically solved using the Hamiltonian method or Pontryagin's Maximum Principle[32]. The optimization problem is defined as:

$$\max_{Y(t)} \int_0^\infty e^{-\delta t} \pi(B(t), Y(t)) dt$$

Subject to the biological state equation:

$$\dot{B} = F(B) - Y(t)$$

where $\pi(\cdot)$ represents the instantaneous profit function and $\delta$ denotes the social discount rate[33]. The solution to this system yields the "Fundamental Equation of Renewable Resources", which establishes an equilibrium condition equating the resource's internal rate of return (marginal biological productivity plus marginal stock effects on harvesting costs) to the external discount rate[2,34,35].

A pivotal insight from dynamic bioeconomics is the sensitivity of the optimal steady-state stock ($B^*$) to the discount rate $\delta$. In the limit where $\delta \to 0$ (implying equal valuation of future and present), the optimal strategy converges to the Maximum Economic Yield (MEY)[36,37]. Conversely, as $\delta \to \infty$, the discounting of future benefits intensifies, driving the optimal solution toward the Open Access Equilibrium (OAE). This implies that if the financial return on alternative assets exceeds the biological growth rate of the stock, a rational sole owner may find it optimal to liquidate the resource entirely. Consequently, resource degradation is not exclusively a failure of property rights (the "Tragedy of the Commons") but can also arise from intertemporal arbitrage between biological and financial capital[38].

## 3. Game Theoretic Foundations

The deterministic ODE models assume a single "social planner" or an aggregate of identical agents acting in unison[39]. However, the tragedy of the commons is fundamentally a strategic problem: the outcome depends on the interdependent decisions of multiple autonomous agents[40]. Game theory provides the lexicon to analyze these strategic interactions.

### 3.1 Prisoner's Dilemma and Strategic Interaction

The archetypal formalism for analyzing the conflict between individual rationality and collective sustainability in Common-Pool Resources is the Prisoner's Dilemma (PD). Considered within a dyadic framework, agents maximize utility by selecting between two discrete strategies: Cooperation ($C$), representing restraint in harvesting, and Defection ($D$), representing intensive extraction.

The structural definition of the dilemma relies on a strict inequality of payoffs: $T > R > P > S$, where: $T$ (Temptation) represents the payoff for unilaterally defecting (free-riding). $R$ (Reward) represents the payoff for mutual cooperation. $P$ (Punishment) represents the payoff for mutual defection (resource depletion). $S$ (Sucker's Payoff) represents the loss incurred by cooperating while the other defects. Additionally, the condition $2R > T + S$ is typically imposed to ensure that mutual cooperation yields the global maximum for collective utility[41].

Analysis of this strategic environment reveals that Defection is a strictly dominant strategy. An agent increases their utility by defecting regardless of the opponent's choice (since $T > R$ and $P > S$). Consequently, the unique Nash Equilibrium of the one-shot game is mutual defection $(D, D)$. This outcome is Pareto-inefficient, as the equilibrium payoff $(P, P)$ is strictly inferior to the cooperative outcome $(R, R)$. This mathematical demonstration of suboptimal equilibrium provided the theoretical underpinning for early bioeconomic governance, reinforcing the "Tragedy of the Commons" narrative and justifying policy dichotomies that favored either privatization or centralized "Leviathan" state control[42,43].

## 3.2 N-Person Commons Game and Group Size Effects

Extending the analysis beyond dyadic interactions, the N-person game generalizes the tragedy of the commons to polyadic settings, where complexity scales with group size[44,45]. The fundamental dilemma is captured by the payoff function for an individual agent $i$ within a group of size $N$, extracting resource quantity $x_i$:

$$\pi_i(x_i, \mathbf{x}_{-i}) = x_i \cdot AP\left(\sum_{j=1}^{N} x_j\right) - c \cdot x_i$$

where $\mathbf{x}_{-i}$ denotes the strategies of all other agents, $c$ represents the constant private marginal cost of effort, and $AP(X)$ is the Average Product (value per unit of effort), which is a monotonically decreasing function of aggregate extraction $X = \sum x_j$.

The core structural failure arises from the asymmetry in cost distribution. While the marginal benefit of extraction accrues privately to the individual, the marginal cost associated with stock depletion (the degradation of $AP$) is distributed across all $N$ participants. Specifically, the individual internalizes only a fraction $1/N$ of the negative externality they generate. As $N \to \infty$, the internalized cost of degradation approaches zero, thereby diminishing the incentive for restraint and increasing the propensity for free-riding[46].

This dynamic leads to a Nash Equilibrium aggregate extraction ($X_{NE}$) that significantly exceeds the socially optimal extraction level ($X_{opt}$). To quantify this inefficiency, algorithmic game theory employs the "Price of Anarchy" (PoA), defined as the ratio between the social welfare at the optimal state and the social welfare at the worst-case Nash equilibrium. In the context of the N-person commons, the PoA typically worsens as $N$ increases, providing a rigorous mathematical validation for the difficulty of achieving cooperative governance in large-scale systems[47,48].

## 3.3 Dynamic Games and Folk Theorem

The deterministic inevitability of resource dissipation described in static models is fundamentally relaxed when the temporal dimension is extended to repeated interactions[49]. Real-world common-pool resource usage is rarely a one-shot event; rather, it constitutes a "supergame" or infinitely repeated game. Within this framework, the Folk Theorem asserts that any feasible, individually rational payoff profile (where outcomes exceed the minimax reservation value $P$) can be supported as a Nash Equilibrium, provided that the discount factor $\delta$ is sufficiently high[50].

Cooperative equilibria are sustained through contingent strategies that link current behavior to future consequences, such as Grim Trigger (permanent reversion to defection following a single breach) or Tit-for-Tat (reciprocity)[51]. For the Grim Trigger strategy to sustain cooperation, the present value of the cooperative stream must exceed the immediate gain from defection followed by the discounted stream of punishment: $\frac{R}{1-\delta} \geq T + \frac{\delta P}{1-\delta}$. Rearranging this inequality yields the critical threshold for $\gamma \in [0,1]$ the discount factor:

$$\delta \geq \frac{T-R}{T-P}$$

This condition formalizes the intuition that cooperation is viable only when the "shadow of the future" is sufficiently long—that is, when agents value future earnings enough that the threat of infinite punishment outweighs the temptation of immediate free-riding[52].

However, while theoretically sound, "unforgiving" strategies like Grim Trigger are practically brittle. Theoretical refinements highlight their vulnerability to imperfect monitoring and environmental noise. In stochastic CPR systems, a natural fluctuation in resource yield could be misidentified as a defection, triggering a permanent collapse into the non-cooperative state $(P, P)$. This fragility has motivated research into more robust mechanisms, such as Graduated Sanctions, which allow for error correction and re-stabilization without precipitating total system failure[53-56].

## 4. Institutional Analysis and Design

While classical game theory predicted tragedy in the absence of external force, Elinor Ostrom's extensive empirical work demonstrated that many communities successfully manage CPRs through self-governance[57].

This contradiction between theory and reality led to the development of the Institutional Analysis and Design (IAD) framework, which has been increasingly formalized mathematically to capture the nuance of "institutions" (rules of the game)[58].

## 4.1 Formalizing Design Principles

Moving beyond the "governance-free" assumptions of the standard Prisoner's Dilemma, Ostrom's empirical framework identifies eight design principles characteristic of robust CPR institutions, including clearly defined boundaries, effective monitoring, graduated sanctions, and low-cost conflict resolution mechanisms[59,60]. Contemporary bioeconomic modeling seeks to formalize these qualitative insights by explicitly modifying the game's payoff structure or sequential architecture to endogenize these institutional variables.

A primary example is the mathematical operationalization of monitoring. This is typically achieved by introducing a third strategy (e.g., Inspect) or structuring the interaction as a multi-stage inspection game. This modification transitions the system from a deterministic payoff structure to a probabilistic one, where the returns on non-compliance are conditional upon enforcement efficacy[59].

Let represent the probability of detection, which is functionally dependent on the monitoring intensity within the population. If $S$ denotes the magnitude of the sanction imposed upon detection, the expected utility of defection ($E[\pi_D]$) is derived as: $E[\pi_D] = (1-\gamma)T + \gamma(T-S)$. The condition for institutional stability requires that the expected return from defection be suppressed below the payoff for mutual cooperation ($R$). Therefore, cooperation becomes the rational dominant strategy if:

$$\gamma S > T - R$$

This inequality demonstrates that resource stabilization is not solely a function of severe punishment ($S$), but is equally sensitive to the certainty of detection ($\gamma$). This formalization provides the theoretical basis for Ostrom's "graduated sanctions" principle, suggesting that high monitoring frequency allows for lower, socially less costly sanctions while maintaining the incentive compatibility of cooperation[61].

## 4.2 Weissing and Ostrom Irrigation Game

A rigorous formalization of enforcement dynamics in asymmetric systems is provided by the irrigation game developed by Weissing and Ostrom[62]. This model examines the strategic interaction between an upstream "Turn-taker" (TT), who possesses an opportunity for illicit extraction, and a downstream "Turn-waiter" (TW), who bears the burden of monitoring.

The game is structured around discrete strategy sets: the TT chooses between Steal ($S$) and Not Steal ($\neg S$), while the TW selects between Monitor ($M$) and Not Monitor ($\neg M$). The payoff structure is parameterized as follows: $B$, the marginal benefit of water theft to the TT. $L$, the consequential loss of water volume incurred by the TW (negative externality)[63,64]. $C$, the private marginal cost of monitoring effort for the TW. $P$, the sanction (fine) levied against the TT upon detection. $R$, the reward accrued by the TW upon successful detection (representing the value of recovered water or a bounty). $\mu$, the conditional probability of detection given that monitoring occurs. The interaction generates a complex incentive landscape. For instance, in the event of detection (Strategy profile $\{S, M\}$), the TT incurs a penalty (payoff $\approx B - P$ or $-P$), while the TW nets $-L - C + R$. Conversely, unmonitored theft yields maximal payoff for the TT ($B$) and maximal loss for the TW ($-L$).

A central insight from Weissing and Ostrom's analysis is the general absence of a stable pure-strategy equilibrium. Instead, the system typically converges to a Mixed-Strategy Nash Equilibrium (MSNE), where agents randomize their actions with probabilities $p_s^*$ (rate of stealing) and $p_m^*$ (rate of monitoring). This equilibrium reveals a counter-intuitive comparative static: increasing the severity of the fine ($P$) does not necessarily yield a linear reduction in non-compliance[65]. According to the indifference principle of MSNE, a higher $P$ requires the monitor to reduce their frequency of monitoring to keep the thief indifferent. Consequently, severe sanctions can inadvertently degrade the monitoring incentive (since monitoring becomes less profitable as theft creates scarcity), creating a feedback loop that may restore the incentive for illicit extraction.

Theoretically, this framework provides a precise formulation of the "second-order public good dilemma." It questions why a rational agent would incur a private cost ($C$) to provide a public good (rule enforcement)[66,67]. The model demonstrates that in dyadic settings, enforcement relies on direct private capture of benefits ($R$ or the prevention of $L$). In larger $N$-person systems where these private benefits are diluted, the incentive to monitor collapses, necessitating the invocation of meta-norms, social esteem mechanisms, or "pool punishment" institutions to sustain cooperation[68,69].

## 4.3 Sanctioning Mechanisms

A salient divergence between classical non-cooperative game theory and empirical institutional analysis lies in the temporal structure of punishment[70]. While classical "supergame" equilibria often rely on the Grim Trigger strategy—mandating a permanent, infinite-horizon reversion to the non-cooperative Nash equilibrium following a single deviation—Ostrom's fifth design principle advocates for graduated sanctions[71,72].

Formally, graduated sanctions are implemented within dynamic games by introducing state variables that track an agent's history of non-compliance (often conceptualized as a "reputation score"). The sanction magnitude $S_t$ at time $t$ is defined as a monotonically increasing function of cumulative deviations:

$$S_t = f\left(\sum_{k=0}^{t-1} \mathbb{I}(D_k)\right)$$

where $\mathbb{I}(D_k)$ is an indicator function for defection at time $k$.

The theoretical superiority of this approach stems from its robustness to stochasticity and imperfect monitoring (noise). In systems prone to "trembling hand" errors or observational inaccuracies, the brittle nature of Grim Trigger precipitates an irreversible collapse of cooperation due to a single false positive. Conversely, graduated sanctions provide a mechanism for error correction and re-stabilization. Initial low-cost penalties function primarily as informational signals or warnings, allowing agents to correct behavior without triggering the catastrophic dissolution of the cooperative equilibrium. This theoretical resilience aligns with empirical findings that robust CPR institutions prioritize restorative compliance and educational signaling over immediate, maximal deterrence.

## 5. Evolutionary Game Theory

While classical game theory assumes hyper-rational agents calculating Nash Equilibria, Evolutionary Game Theory (EGT) assumes agents adapt their strategies over time based on relative success (payoff). This approach is widely used in modern HES models to describe the "human" side of the coupled system[73,74]. The core question shifts from "what is the equilibrium?" to "how does the frequency of cooperators $x(t)$ evolve over time?" Three primary stochastic processes are used to model this evolution: Replicator Dynamics, the Moran Process, and the Fermi Process. Understanding the specific mathematical mechanics of each is crucial for interpreting model outcomes.

### 5.1 Replicator Dynamics

The fundamental formalism governing the temporal evolution of strategy frequencies in EGT is the Replicator Equation[75,76]. Consider a large, finite population composed of agents adopting one of $n$ distinct phenotypic strategies. Let $x_i$ denote the frequency of strategy $i$ within the population state vector $\mathbf{x}$. The evolutionary trajectory of strategy $i$ is driven by the differential between its specific fitness $f_i(\mathbf{x})$ and the population-wide average fitness $\bar{f}(\mathbf{x})$:

$$\dot{x}_i = x_i[f_i(\mathbf{x}) - \bar{f}(\mathbf{x})]$$

where the average fitness is defined as $\bar{f}(\mathbf{x}) = \sum_{j=1}^{n} x_j f_j(\mathbf{x})$.

In the specific context of Common-Pool Resources, the analysis typically dichotomizes the population into Cooperators and Defectors[77,78]. In a panmictic (well-mixed) population lacking structural constraints, Defectors systematically accrue a fitness advantage via free-riding ($f_D > f_C$)[79]. Consequently, the replicator dynamic dictates a positive growth rate for the defecting phenotype ($\dot{x}_D > 0$), driving the system inexorably toward a monomorphic equilibrium of universal defection ($x_D \to 1$).

This result rigorously corroborates the "Tragedy of the Commons" as the unique basin of attraction in unstructured systems: cooperative traits are driven to extinction by natural selection. However, the EGT framework facilitates a more nuanced analysis of invadability and Evolutionarily Stable Strategies (ESS). While classical models predict that a population of Defectors cannot be invaded by a rare cooperative mutant (making Defection an ESS), theoretical extensions demonstrate that mechanisms such as positive assortativity—where Cooperators preferentially interact with other Cooperators—or explicit spatial structure can modify the fitness landscape, thereby stabilizing cooperative equilibria against invasion[80,81].

## 5.2 Spatial Games and Local Interactions

While standard evolutionary models assume panmixia (well-mixed populations), empirical resource systems possess explicit spatial dimensions (e.g., coastal fisheries, forest grids). To capture this, EGT is frequently mapped onto lattice structures or complex networks, where interactions are restricted to a local neighborhood defined by network topology[82].

In these spatially explicit models, the evolutionary dynamic shifts from global replication to local imitation. Agents update their strategies by adopting the phenotype of the highest-performing neighbor within their interaction radius. This localization fundamentally alters the evolutionary outcome through the mechanism of spatial reciprocity[83].

Specifically, cooperative agents can self-organize into compact clusters. Individuals located within the core of such clusters interact exclusively with other cooperators, thereby accruing the maximal mutual benefit ($R$). This topological configuration effectively "shields" them from the sucker's payoff ($S$), which is incurred only by agents at the cluster periphery in contact with defectors. If the cumulative benefit of the cluster core outweighs the peripheral losses to exploitation ($T$), cooperative regions can persist or expand against the defector invasion.

Mathematical analysis of the Spatial Prisoner's Dilemma reveals a rich spectrum of emergent spatiotemporal dynamics. Contingent upon the specific payoff parameters ($T, R, P, S$), the system may undergo phase transitions between distinct regimes: Frozen States, the formation of stable, static domains of cooperators and defectors. Chaotic Dynamics, a state of persistent, non-repeating spatiotemporal fluctuations. Traveling Waves, cyclical patterns where defectors exploit and deplete cooperative clusters, which subsequently recolonize the resulting voids, mimicking predator-prey dynamics[84].

## 5.3 Quantum Game Theory Approaches

A nascent yet profound theoretical frontier involves the application of Quantum Game Theory (QGT) to common-pool resource dilemmas[85]. This framework transcends the limitations of classical game theory by relaxing the assumption of strategy independence, utilizing quantum mechanical formalism—specifically superposition and entanglement—to model non-trivial correlation structures between agents[86,87].

In the standard Eisert-Wilkens-Lewenstein (EWL) protocol, the strategic interaction is mapped onto a complex Hilbert space rather than a discrete decision set[88]. The evolution of the system state is governed by unitary operations. If $|\psi_{initial}\rangle$ represents the initial state (typically $|00\rangle$), the final state $|\psi_{final}\rangle$ is derived via:

$$|\psi_{final}\rangle = \hat{J}^{\dagger}(\hat{U}_A \otimes \hat{U}_B)\hat{J}|\psi_{initial}\rangle$$

where $\hat{U}_A$ and $\hat{U}_B$ denote the local unitary strategy operators chosen by the players, and $\hat{J}$ represents the entangling operator which introduces correlation prior to the strategic choice.

The pivotal result in this domain is the topological resolution of the Prisoner's Dilemma. Research demonstrates that when the degree of entanglement (parameterized by $\gamma \in [0, \pi/2]$ in the operator $\hat{J}$) exceeds a critical threshold, the classical Nash Equilibrium of mutual defection is destabilized. The expanded Hilbert space allows for "superposed" strategies that are inaccessible in the classical domain, rendering the Pareto-optimal outcome $(R, R)$ a unique Nash Equilibrium.

While currently an abstract formalism, QGT offers a sophisticated mathematical syntax for describing systems defined by extreme interdependence. It serves as a phenomenological analogue for "entangled" social states, providing a rigorous way to model how deep cultural norms, psychological contracts, or binding institutions can correlate agent behaviors to such a degree that independent defection becomes a stochastic impossibility.

# 6. Beyond Homo Economicus

The "rational actor" assumption of standard bioeconomics and classical game theory often fails to match experimental data. Humans demonstrate bounded rationality, social preferences, and cognitive biases[89]. Integrating these into mathematical models has improved their predictive power[90,91].

## 6.1 Prospect Theory in Common Pool Resources

While classical bioeconomics relies on Expected Utility Theory (EUT), recent literature increasingly incorporates Prospect Theory (PT), established by Kahneman and Tversky[92,93], to account for cognitive biases in decision-making. The central tenet of PT is reference dependence: agents evaluate outcomes not as final states of wealth, but as deviations from a reference point, exhibiting asymmetric sensitivity to gains and losses[94,95].

The subjective utility of an outcome $x$ is formalized by the piecewise value function $v(x)$:

$$v(x) = \begin{cases} x^\alpha & \text{if } x \geq 0 \quad \text{(Gains)} \\ -\lambda(-x)^\beta & \text{if } x < 0 \quad \text{(Losses)} \end{cases}$$

where the parameters $\alpha, \beta < 1$ capture the principle of diminishing sensitivity (implying concavity in the gain domain and convexity in the loss domain). The parameter $\lambda > 1$ represents the coefficient of loss aversion, empirically estimated at $\lambda \approx 2.25$, indicating that the marginal disutility of a loss significantly outweighs the marginal utility of an equivalent gain.

In the context of Common-Pool Resources, the "reference point" is typically anchored to the historical status quo of high consumption. Consequently, conservation policies necessitating catch reductions are encoded psychologically as "sure losses." Because the value function is convex in the negative domain ($v''(x) > 0$ for $x < 0$), agents exhibit risk-seeking behavior when facing losses. Confronted with a choice between a deterministic reduction in income (regulation) and a probabilistic outcome involving either maintaining high yield or total stock collapse, loss-averse agents irrationally prefer the risky gamble. This phenomenon, often termed "gambling for resurrection," rationalizes the resistance to effort reduction observed in declining fisheries. Theoretical comparisons by Sundaram et al[96]. demonstrate that this behavioral asymmetry shifts the Nash equilibrium, significantly increasing the probability of resource exhaustion compared to predictions based on standard neutral utility.

## 6.2 Hyperbolic Discounting

A critical deviation from standard bioeconomic theory involves the modeling of intertemporal choice. While classical models assume a constant discount rate (exponential discounting), empirical evidence from behavioral economics suggests that human time preference is frequently hyperbolic[97,98]. In this framework, the discount factor $D(t)$ is more accurately represented as:

$$D(t) = \frac{1}{1+kt}$$

standing in contrast to the standard exponential formulation $D(t) = e^{-\delta t}$.

This functional form generates dynamic inconsistency (or preference reversals). Mathematically, the discount rate is not constant but declines as the time horizon extends. Consequently, an agent at $t=0$ may rationally prefer a conservation strategy for a future date $t=\tau$. However, as time advances and $t \to \tau$, the "immediate" utility of consumption is heavily weighted due to present bias, causing the agent to reverse their preference and choose extraction.

This mechanism offers a robust behavioral explanation for the prevalence of non-compliance in long-term governance frameworks (e.g., climate accords), where actors commit to future sustainability goals but defect in the short term[99]. Consequently, robust institutional design requires commitment devices—binding mechanisms that constrain the choice set of future selves to enforce the original optimal plan. Furthermore, theoretical models involving "naive" agents (who fail to anticipate their future preference shifts) demonstrate that hyperbolic discounting significantly lowers the threshold for resource collapse, particularly in species with slow intrinsic growth rates where the delay between conservation cost and biological recovery is significant[100].

# 7. Stochastic Dynamics, Nonlinearity, and Tipping Points

Contemporary modeling frameworks increasingly conceptualize Common-Pool Resources as Coupled Human-Environment Systems[101-103]. Central to this paradigm is the recognition of reciprocal feedback mechanisms linking social and ecological subsystems. This dynamic interplay frequently engenders complex system behaviors, characterized by high nonlinearity, the emergence of alternative stable states, and a heightened vulnerability to abrupt systemic collapse.

## 7.1 Stochastic Differential Equations and Environmental Noise

To capture the inherent aleatory uncertainty of natural systems—driven by fluctuations in environmental covariates such as precipitation or temperature—deterministic paradigms are frequently augmented using Stochastic Differential Equations (SDEs). The canonical logistic growth ODE is typically transformed into a geometric Brownian motion framework by introducing a stochastic forcing term. The resulting dynamics are governed by the Itô SDE:

$$dX_t = rX_t\left(1 - \frac{X_t}{K}\right)dt + \sigma X_t dW_t$$

where $dW_t$ represents the increment of a standard Wiener process (Brownian motion), and $\sigma$ denotes the coefficient of environmental volatility (noise intensity).

This transition from deterministic to stochastic modeling fundamentally alters the definition of system stability. The concept of a fixed point equilibrium ($B_{eq}$) becomes ill-defined; instead, the system's long-term behavior is characterized by a stationary probability density function, often derived via the Fokker-Planck equation. Consequently, management objectives must be reframed from maintaining a specific biomass target to risk-based metrics, specifically minimizing the probability of extinction or maximizing the Mean First Passage Time—the expected duration before the stock trajectory crosses a critical quasi-extinction threshold[104-106].

## 7.2 Allee Effect and Critical Depensation

A pivotal source of nonlinearity in population dynamics is the Allee effect, characterized by a positive correlation between population density and per-capita fitness at low abundance levels (depensation). Unlike standard logistic growth, where low density promotes rapid recovery, a "strong" Allee effect introduces an unstable equilibrium point or critical depensation threshold ($A$)[107,108]. If the biomass trajectory breaches this lower bound ($X < A$), the growth rate becomes negative, precipitating a deterministic spiral toward extinction[109-111].

This dynamic is formally captured by modifying the logistic equation to include a threshold term:

$$\frac{dX}{dt} = rX\left(1 - \frac{X}{K}\right)\left(\frac{X}{A} - 1\right) - Y$$

where the term $(\frac{X}{A} - 1)$ enforces negative growth when $X < A$.

When coupled with stochastic environmental forcing ($dW_t$), the presence of an Allee threshold fundamentally alters the system's risk profile. Even when the system resides at a deterministically stable equilibrium above $A$, stochastic fluctuations introduce a non-zero probability of the state variable crossing the separatrix (the unstable manifold at $A$) into the basin of attraction for extinction. Empirical analyses suggest that the omission of this mechanism from standard MSY calculations constitutes a primary driver of fishery collapse. Stocks managed under simple logistic assumptions may be harvested down to levels that appear sustainable deterministically, but which lack the resilience to withstand stochastic shocks in the presence of latent depensatory feedbacks[112,113].

## 7.3 Tipping Points and Early Warning Signals

A central trajectory of contemporary research concerns the anticipation of catastrophic bifurcations—tipping points where a system undergoes a sudden, often irreversible transition from a desirable regime to a degraded state (e.g., fishery collapse)[114]. The theoretical basis for detecting such shifts relies on the phenomenon of Critical Slowing Down (CSD), which occurs as a system approaches a local bifurcation point (specifically a fold or Hopf bifurcation)[115,116].

Dynamically, CSD arises as the real part of the dominant eigenvalue ($\lambda$) of the system's linearized Jacobian matrix approaches zero[117,118]. In topological terms, this corresponds to a flattening of the potential landscape (the stability basin becomes shallower), implying a progressive weakening of the restoring forces that normally return the system to equilibrium following a perturbation. Mathematically, this loss of resilience leaves generic statistical fingerprints in the time-series data, serving as Early Warning Signals (see Fig. 2).

As the restoring rates vanish ($\lambda \to 0$), the system preserves the "memory" of perturbations for longer durations. This manifests as an increase in the lag-1 autocorrelation coefficient ($AR(1)$)[119,120]. Modeling the deviations as a discrete-time autoregressive process:

$$x_{t+1} = \alpha x_t + \epsilon_t$$

The autoregressive coefficient $\alpha$ corresponds to $e^{\lambda \Delta t}$. Thus, as the system approaches criticality ($\lambda \to 0$), $\alpha \to 1$.

The flattening of the potential well implies that even constant levels of environmental noise result in larger stochastic excursions from the equilibrium. The variance of the state variable scales inversely with the eigenvalue magnitude:

$$Var(x) \propto \frac{\sigma^2}{|\lambda|}$$

Consequently, as $|\lambda| \to 0$, the variance theoretically diverges ($Var \to \infty$), a phenomenon often termed "flickering" or noise amplification prior to transition[115,121].

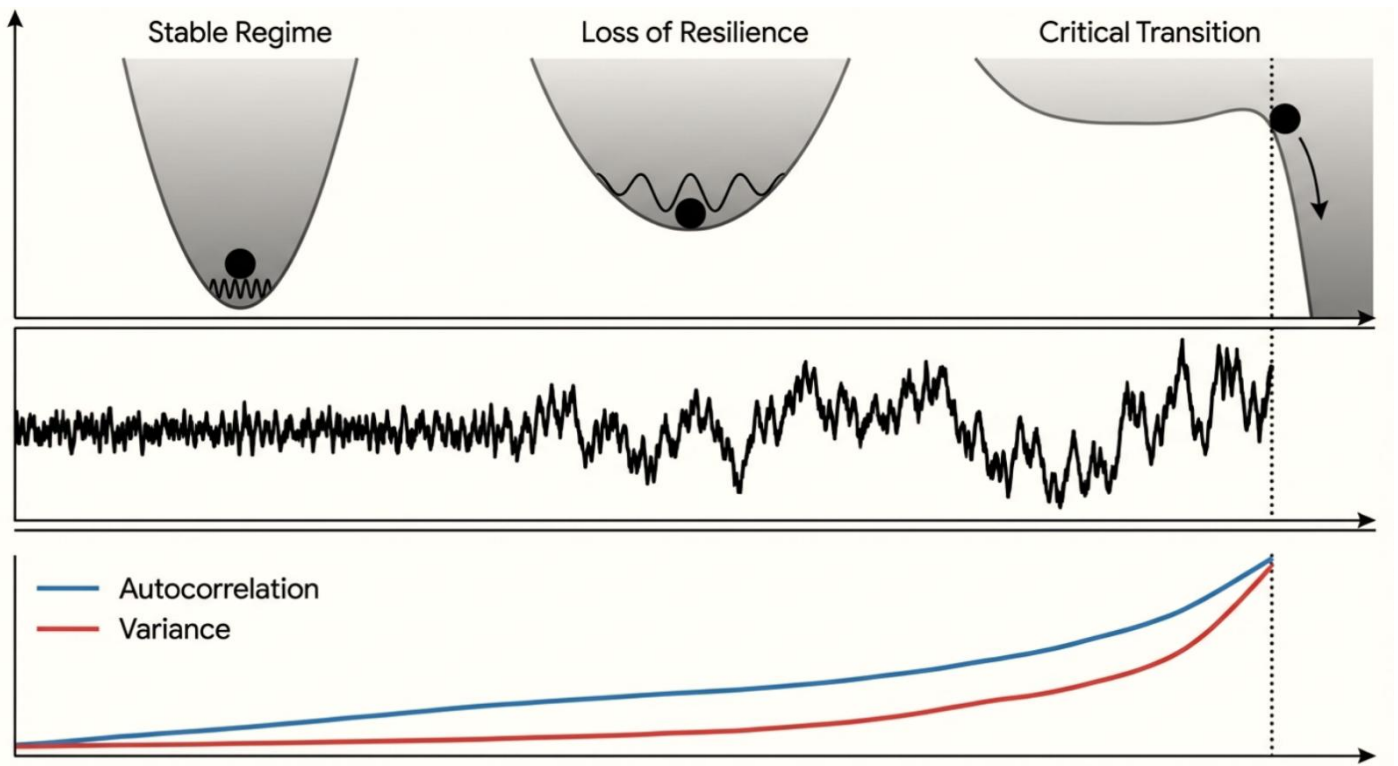

**Figure 2. The Anatomy of a Tipping Point and Critical Slowing Down.** This schematic illustrates the loss of resilience as a system approaches a critical transition. The stability landscape, represented as a potential well, becomes progressively shallower (flatter), indicating a loss of restoring force. As the well flattens, the system state (depicted as a ball) experiences 'critical slowing down,' taking longer to return to equilibrium after stochastic perturbations. This results in larger excursions from the mean. Corresponding early warning signals (EWS) derived from the time series data. Both variance and lag-1 autocorrelation exhibit a marked increase as the system nears the tipping point, preceding the eventual collapse.

## 8. Agent-Based Modeling and Computational Approaches

While analytical models (ODE, Game Theory) provide elegance and tractability, they often require simplifying assumptions (homogeneity, perfect mixing, infinite rationality) that limit their realism. Agent-Based Models (ABMs) allow researchers to simulate the interactions of heterogeneous agents in explicit spatial environments, creating a bottom-up generative science of CPRs.

## 8.1 Heterogeneity and Spatiality

A primary analytical advantage of Agent-Based Models (ABMs) is their capacity to parameterize multidimensional heterogeneity across agent populations, moving beyond the "representative agent" constraint[122]. Simulations can explicitly encode variance in harvesting capabilities, initial endowments, and cognitive beliefs. While foundational literature suggests that economic inequality frequently destabilizes cooperative equilibria, computational experiments reveal that this relationship is topologically contingent: in specific network structures, agents with superior endowments may function as high-centrality "hubs," effectively subsidizing enforcement costs and anchoring compliance norms within the system.

Furthermore, spatially explicit ABMs facilitate the rigorous investigation of bounded rationality through restricted observational horizons. Unlike canonical game-theoretic models that often assume global information availability, ABM agents operate under "imperfect information," where visibility is constrained to a local spatial radius. This localization fundamentally restructures the emergent dynamics. Counter-intuitively, empirical and computational research indicates that information restriction can promote sustainability. Localized perception fosters the self-organization of "trust clusters" or protected niches of reciprocity. Conversely, global information transparency can inadvertently catalyze a "contagion of defection," where agents preemptively react to distant non-compliance, triggering a systemic collapse or "race to the bottom" that would be absent under strictly local interaction rules[123,124].

## 8.2 The ADICO Grammar in Code

A significant methodological advancement in computational social science is the direct integration of Crawford and Ostrom's "Grammar of Institutions" into the architecture of Agent-Based Models[125]. This framework facilitates the rigorous syntactic decomposition of institutional statements into five discrete components, rendering social rules computationally tractable (see Fig. 3)[126]:

- Attributes ($A$): The subset of the population to whom the statement applies (e.g., role-specific targeting).
- Deontic ($D$): The modal operator prescribing the nature of the constraint (e.g., permissible, obligatory, forbidden).
- Aim ($I$): The specific action or outcome regulated by the rule.
- Conditions ($C$): The spatial, temporal, or state-based parameters under which the rule is active.
- Or else ($O$): The punitive consequence associated with non-compliance, distinguishing binding rules from mere norms.

By encoding these components as mutable parameters rather than static constraints, researchers can simulate institutional evolution. This approach allows for the endogenization of governance: agents can algorithmically "propose" syntactical mutations (e.g., shifting a parameter from "May" to "Must Not") and engage in collective choice mechanisms based on historical utility maximization. Computational experiments demonstrate that such systems can endogenously evolve robust, self-organized regulatory regimes—such as rotation schemes or quota systems—in the absence of exogenous imposition[127]. However, the stability of these emergent institutions is frequently shown to be contingent upon the fidelity of environmental feedback, degrading when the signal-to-noise ratio in the resource system obscures the causal link between rule compliance and payoff.

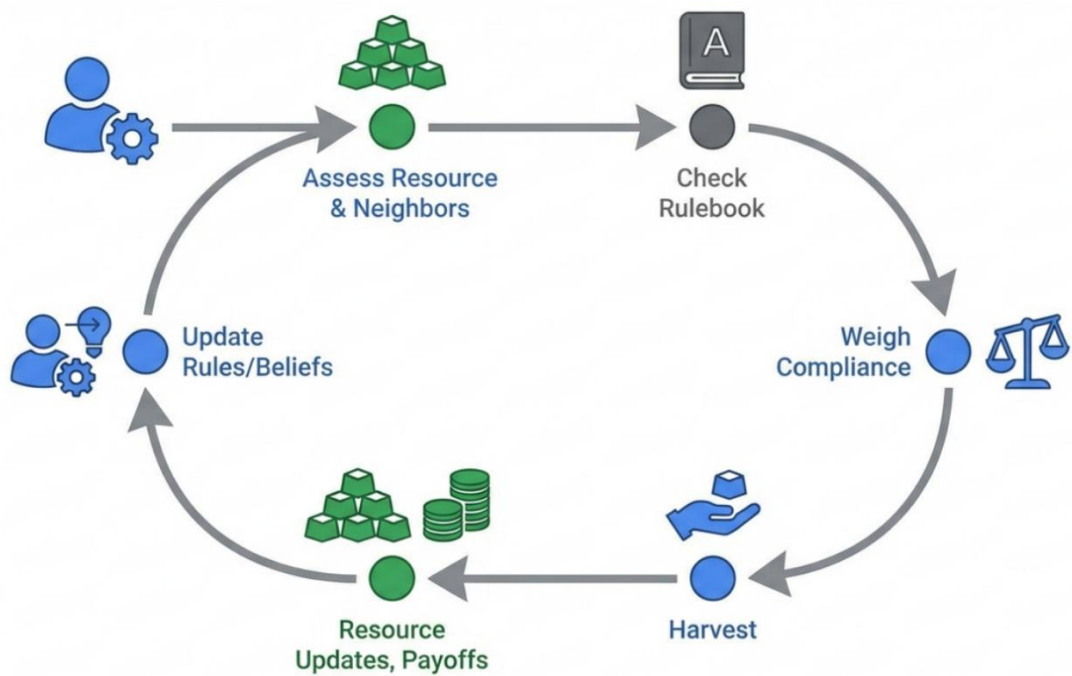

**Figure 3: The Agent Logic Loop: Integrating Institutional Grammar into Micro-Level Decision Making.** This schematic details the cognitive and behavioral cycle of a single agent within one simulation time-step of the Common-Pool Resource model: 1. Perception: Agent assesses resource state and neighbor behavior. 2. Institutional Filter: Agent checks internal rulebook (e.g., 'Must not harvest > 5 units'). 3. Strategy Selection: Agent weighs compliance vs. cheating (payoff expected). 4. Action: Harvest occurs. 5. Environmental Feedback: Resource stock updates; Payoffs distributed. 6. Learning: Agent updates rules/beliefs based on outcome.

## 8.3 Coupled Dynamics and Feedback Sensitivity

The defining architectural feature of Coupled Human-Environment System models is the explicit structural integration of biophysical state variables with socio-economic decision algorithms[128]. This co-evolutionary framework is mathematically formalized as a system of coupled differential equations:

$$\frac{dR}{dt} = f(R, E)$$

$$\frac{dE}{dt} = g(E, R, \text{Social Information})$$

where the ecological dynamics ($f$) are directly conditioned by harvesting effort ($E$), while the evolution of effort ($g$) is a function of both the resource state ($R$) and a vector of social information ($\Omega$). The function $g(\cdot)$ typically operationalizes adaptive social learning, where agents modulate their strategies based on payoff-biased imitation or the perceived abundance of the stock.

A critical dynamical property emerging from this interaction is the sensitivity to coupling strength—specifically, the relative timescales of social adaptation versus biological regeneration. When the social system exhibits high reactivity (tight coupling)—meaning fishing effort responds instantaneously to transient fluctuations in biomass—the system is prone to destabilization. This lack of dampening frequently results in overcompensation, driving the trajectory away from stable equilibria and into high-amplitude boom-bust regimes (limit cycles). For instance, if harvesting pressure scales too aggressively with marginal increases in stock, the system may overshoot the sustainable yield threshold, precipitating a population crash before biological compensatory mechanisms can restore stability.

# 9. Synthesis and Future Directions

The mathematical modeling of common-pool resources has traversed a significant intellectual arc. It began with the Gordon-Schaefer recognition that open access dissipates rent. It moved to Game Theory, which diagnosed the strategic tragedy and the potential for repeated-game cooperation. Ostrom's IAD framework added institutional texture, which was formalized through Monitoring Games. Finally, Coupled Human-Environment System and Agent-Based Model approaches have acknowledged that these systems are nonlinear, stochastic, and evolving.

## 9.1 Key Insights from the Mathematical Synthesis

The mathematical trajectory of Common-Pool Resource modeling—from the deterministic Gordon-Schaefer model to complex adaptive systems—yields several critical theoretical insights that challenge foundational bioeconomic assumptions:

1. The Insufficiency of Static Equilibrium: Classical static analysis systematically underestimates systemic vulnerability by neglecting transient dynamics and intertemporal inconsistencies. The incorporation of hyperbolic discounting reveals that even ostensibly optimal equilibrium paths are frequently time-inconsistent, leading to defect-driven collapse during the transition phase[46].

2. The Structural Fragility of Cooperation: While the Folk Theorem establishes the theoretical possibility of cooperative Nash Equilibria in repeated games, these solutions exhibit structural instability. In the presence of stochastic noise and agent heterogeneity, "grim trigger" mechanisms fail. Robust cooperation necessitates second-order enforcement mechanisms (punishment), the provision of which constitutes a "second-order social dilemma" often unresolved in pure self-interest models[129,130].

3. Behavioral Asymmetry as a Driver of Collapse: The integration of Prospect Theory identifies psychological asymmetry as a destabilizing force. Loss aversion anchors agents to historical consumption levels, transforming necessary conservation efforts into perceived losses. This convex value function induces risk-seeking behavior ("gambling for resurrection"), thereby accelerating depletion rates precisely when stock conservation is most critical[40,131].

4. The Stabilizing Role of Topology: Contrary to the pessimistic predictions of mean-field (panmictic) mixing models, Agent-Based and Network approaches demonstrate that spatial localization promotes sustainability. Explicit topological structure facilitates the formation of "trust clusters," where local reciprocity shields cooperators from global defection pressures[132].

5. From Optimization to Resilience: The shift toward Stochastic Differential Equations (SDEs) and bifurcation analysis necessitates a paradigmatic transition in management objectives. In a stochastic environment characterized by non-linear tipping points, the precise maximization of yield (MSY) is increasingly viewed as a brittle strategy[15,133].

Consequently, the focus of modern bioeconomics has shifted toward resilience engineering—managing the system not to maximize flow at a specific point, but to maximize the width and depth of the basin of attraction, ensuring the system remains within a "safe operating space" despite inevitable environmental and behavioral volatility.

## 9.2 Future Horizons

The next generation of computational inquiry is poised to transcend fixed-rule heuristics through the integration of Deep Reinforcement Learning (DRL) into agent-based architectures[134]. Unlike traditional agents constrained by pre-defined behavioral protocols, DRL agents utilize neural networks to approximate value functions over high-dimensional state spaces. This allows for the endogenous derivation of complex strategic policies—such as rotational harvesting, costly signaling, or deceptive mimicry—that emerge from trial-and-error optimization without a priori specification by the modeler[135,136].

Concurrently, the theoretical domain is expanding to encompass Quantum Game Theory (QGT) as a rigorous formalism for analyzing high-order social correlation[137]. Beyond its mathematical novelty, QGT offers a sophisticated phenomenological analogue for the "entangled" nature of social contracts observed in tightly-knit or indigenous communities, where cultural norms create non-local correlations in decision-making that defy classical assumptions of agent independence[138].

Ultimately, the overarching objective of this mathematical synthesis is to shift the discipline's focus from merely diagnosing the Tragedy of the Commons to the proactive engineering of resilience. By translating qualitative social dilemmas into rigorous formal systems, we acquire the analytical precision necessary to design institutions that structurally align individual utility maximization with the long-term stability of the biosphere.

# Acknowledgements

This paper is supported by Zhejiang Provincial Philosophy and Social Sciences Planning Project (Grant No. 24NDJC175YB). C.T. acknowledges support from the National Natural Science Foundation of China (Grant No. 72571247) and Scientific Research Project of Zhejiang Provincial Bureau of Statistics (Grant No. 25TJZZ18).

# Conflicts of Interest

The authors declare no conflicts of interest.